\documentclass[12pt,preprint]{aastex}

\usepackage{multirow}
\usepackage{graphicx, subfigure}

\shorttitle{Dark Hole via SLM}
\shortauthors{Dou \& Ren}

\begin{document}


\title{Phase Quantization Study of Spatial Light Modulator for Extreme High-contrast Imaging}


\author{Jiangpei Dou\altaffilmark{1,2,3} and Deqing Ren\altaffilmark{1,2,3}}


\altaffiltext{1}{Physics \& Astronomy Department, California State University Northridge, 18111 Nordhoff Street, Northridge, CA 91330, USA; jpdou@niaot.ac.cn, jiangpeidou@gmail.com.}
\altaffiltext{2}{National Astronomical Observatories / Nanjing Institute of Astronomical Optics \& Technology, Chinese Academy of Sciences, Nanjing 210042, China.}
\altaffiltext{3}{Key Laboratory of Astronomical Optics \& Technology, Nanjing Institute of Astronomical Optics \& Technology, Chinese Academy of Sciences, Nanjing 210042, China.}


\begin{abstract}
Direct imaging of exoplanets by reflected starlight is extremely challenging due to the large luminosity ratio to the primary star. Wave-front control is a critical technique to attenuate the speckle noise in order to achieve an extremely high contrast. We present a phase quantization study of a spatial light modulator (SLM) for wave-front control to meet the contrast requirement of detection of a terrestrial planet in the habitable zone of a solar-type star. We perform the numerical simulation by employing the SLM with different phase accuracy and actuator numbers, which are related to the achievable contrast. We use an optimization algorithm to solve the quantization problems that is matched to the controllable phase step of the SLM. Two optical configurations are discussed with the SLM located before and after the coronagraph focal plane mask. The simulation result has constrained the specification for SLM phase accuracy in the above two optical configurations, which gives us a phase accuracy of 0.4/1000 and 1/1000 waves to achieve a contrast of $10^{-10}$. Finally, we have demonstrated that a SLM with more actuators can deliver a competitive contrast performance on the order of $10^{-10}$ in comparison to that by using a deformable mirror.
\end{abstract}





\section{INTRODUCTION}
High contrast imaging of exoplanets has made important progress in recent years (Lagrange et al. 2010; Marois et al. 2010; Macintosh et al. 2015; Wagner et al. 2016). However, most of the imaged planets are detected and characterized via self emission in the infrared wavelength range. To image a planet by reflected starlight in the visible spectrum still remains challenging. The contrast ratio will be $10^{-10}$ for an Earth-size planet in the habitable zone of a solar-type star (Brown \& Burrow 1990). Future giant segmented mirror telescopes, including TMT/PFI (Macintosh et al. 2006) and E-ELT/EPICS (Kasper et al. 2010), have all been proposed to be applied for the direct imaging and spectroscopy research of super-Earth or rocky planets in the habitable zone (Guyon et al. 2012); however, it is still widely believed that the search for terrestrial life signals should be conducted in future space missions (Bord\'{e} \& Traub 2006). In recent years, several space coronagraph conceptual programs have been proposed to image mature exoplanets at short orbital separations down to $0.1\arcsec$ (ACCESS by Trauger et al. 2012; SPICES by Boccaletti et al. 2012; CPI by Dou et al. 2015).

A variety of coronagraphs have been proposed to tackle the photon noise diffracted from the telescope pupil, and can reach a contrast of $10^{-10}$ in theory (Guyon et al. 2006). However, the achievable contrast is around $10^{-7}$ due to the limitation of speckle noises from imperfect optics without wave-front correction. Wave-front control will be critical to eliminate the speckle noise. Malbert et al. (1995) firstly introduced a dark hole algorithm to achieve a contrast on the order of $10^{-8}$ in a local region by using a deformable mirror (DM). Bord\'{e} \& Traub (2006) refined the solution with faster convergence rate and improved the contrast down to $10^{-10}$ in theory by introducing a coronagraph. A contrast of $0.6 \times10^{-9}$ was firstly demonstrated in the lab on the High Contrast Imaging Testbed (HCIT, Trauger \& Traub 2007). Recently, several experiments have conducted on HCIT by using a DM with similar numbers of actuator elements, but optimized for different types of coronagraph, including the Vortex coronagraph (Serabyn et al. 2013), Pupil mapping coronagraph (Kern et al. 2013) and the Shaped pupil coronagraph (Belikov et al. 2007). The advantage of using DM is that it has a very high phase accuracy, which has demonstrated a sub-nm root mean square (rms) residual wave-front error for a MEMS DM (Evens et al. 2006); however, among these test results, the area of the created dark hole is relatively small, which is limited by the number of elements in the DM.

The application of a spatial light modulator (SLM) for high-contrast coronagraph was first proposed by Ren \& Zhu (2011), in which one SLM and DM is used for active pupil apodization and phase corrections, respectively. We then initially demonstrated a contrast on the order of $10^{-5}$ (Zhang et al. 2012) and improved to $10^{-6}$ under a two-SLM configuration, with the second SLM for wave-front control instead of using a DM (Dou et al. 2014). It was found that the SLM with more actuator elements has an advantage when used for phase corrections. Recently, we applied the SLM in our coronagraph system for wave-front control. A contrast of $1.7\times10^{-9}$ has been achieved in a large area (Ren \& Zhu 2007; Liu et al. 2015).

As demonstrated in our previous work, an SLM could potentially be used in extreme high contrast imaging instead of DM. SLM would increase the detection area with more actuator elements. To fully understand the potential of SLMs for extreme high contrast, we present a feasibility study of a SLM in a high-contrast imaging system by using numerical simulation. In the simulation, we will focus on the phase quantization effect and discuss the possibility of improving the phase steps for the current SLM. Other optical limitations to SLMs such as the polarization stability or homogeneity of the actuators will need further analysis and will not be discussed in this paper. We first present the dark hole generation for a square-shaped clear pupil system, without a coronagraph. It is found that there is a contrast limitation on the order of $10^{-8}$, in which the contrast will not be improved by further increasing the actuator numbers. Then we introduce a coronagraph with a precise wave-front control to improve the contrast. We employ two optical configurations with the SLM before and after the coronagraph focal mask. It is found that SLMs with phase step resolution of 0.4/1000 and 1/1000 waves are needed to achieve a contrast down to $10^{-10}$ for the above two configurations, respectively. Finally, we demonstrate that SLMs with more actuators numbers can deliver a competitive contrast performance in comparison to that by using a DM.

In Section 2, we present the optimization algorithm and associated numerical simulation results for a clear pupil system. The layout and simulation results for a stepped-transmission filter coronagraph are presented in Section 3. In Section 4, we discussed the potential limitations and possible improvements of the SLM. Finally, conclusions and future work are presented in Section 5.

\section{HIGH-CONTRAST IMAGING FOR A SQUARE-SHAPE CLEAR PUPIL SYSTEM}
\subsection{Optimization Algorithm}


In this section, we introduce a phase modulation for a square-shape clear pupil, to demonstrate the limitations of SLMs and DMs in the generation of a dark hole, with a certain phase accuracy and number of actuator elements. We also define the inner working angle (IWA) and out working angle (OWA) of the high-contrast imaging system for the following numerical simulation section, which correspond to the start and end edge of the to-be-generated dark hole, respectively.

The point spread function (PSF) of the system is a square of the complex modulus of the phase-modulated electric field on the focal plane, and is given as
\begin{equation}
\ I_{psf} (x,y)=|\digamma [P(u,v)e^{i\varphi(u,v)}|^{2}.
\end{equation}
where $\digamma$ represents the Fourier transform of the associated function;  $P(u,v)$ represents the entrance pupil of the optics system, with a square shape of $N \times N$ pixels; and $\varphi(u,v)$ is the phase to be introduced to generate a dark hole in the focal plane image.

To find an optimal phase to generate a dark hole, the algorithm minimizes the light intensity in a certain region in the PSF, where the introduced phase is a variable of the optimization (Dou et al. 2011; Ren et al. 2012).

For demonstration purposes, we define the dark hole in a square-shaped area in one quarter of the PSF, by using one single SLM or DM in the following simulation. The optimization algorithm is then to minimize the following equation:

\begin{equation}
\ \frac{I_{out} (x,y)}{I_{in} (x,y)},
\end{equation}
 subject to $0\leq \varphi (u,v)\leq\lambda ~~\&~~ \Delta\varphi= G_{i} \varphi_{a}$,

Figure 1 shows the applications of the algorithm. $I_{in}(x,y)$ is the starlight intensity defined within four diffraction beams (IWA=$4 \lambda/D$) distance from the PSF center. And $I_{out}(x,y)$ is the intensity defined in a square area in one quarter of the PSF. The algorithm minimizes the intensity in the area defined by $I_{out}(x,y)$, which can generate a dark hole with a contrast gain in the star PSF image; meanwhile, the algorithm will keep the intensity attenuation as low as possible for the planet, thus the scattered light due to wave-front modulation will be pushed into the central part of the PSF. Due to the Nyquist limit, OWA can only reach N /2 times of $\lambda/D$, when $N \times N$ array actuator elements are used for the phase modulation or wave-front control.

The algorithm will work in iteration mode and the phase variable of $\varphi(u,v)$ will be updated in each iteration step to minimize the objective function. $\Delta \varphi$ is the phase variation of the actuator elements between two neighboring optimization iteration steps. The optimization becomes an integer variable problem, due to the controllable phase step of the actuator elements. Thus, $\Delta \varphi$ is written in the form of $G_{i}\varphi_{a}$, where $G_{i}$ is an integer variable ($i=1, 2, ..., N^{2}$) and $\varphi_{a}$ corresponds to the phase step of the actuator elements. Based on the current technique, the typical phase resolution of the DM (with a stroke of $0.5\sim1.5 \mu m$ and controlled with 14 bit D/A conversion electronics) and the SLM (Meadowlark Optics) can reach $\lambda/10,000$ and $\lambda/1000$, respectively. Therefore, we will set $\varphi_{a}$ to be $\lambda/1000$ and $\lambda/10,000$ for the SLM and DM. The algorithm is to find the optimal integer values of $G_{i}$ to minimize Equation (2). In the simulation, we will employ simulannealbnd, a global optimization solver in MATLAB, to solve the integer variable problems. The solver uses the simulated annealing algorithm to find a global minimum of a function. For details of the algorithm, the reader can refer to the paper of Brooks \& Morgan (1995), or the Global Optimization Toolbox User Guide of MATLAB.

\subsection{Numerical Simulation of Dark Hole Generation for a Clear Pupil}
In recent years, a DM manufactured by Xinetics Inc has been used for dark hole tests in the laboratory, with $32 \times 32$ and $64 \times 64$ actuator numbers. Therefore, we use these two kinds of actuator numbers to conduct the following numerical simulation.

It is found the optimization procedure has a low convergence rate if it is started from a random phase. Therefore, in the first step, we sought the start point of $G_{i}$ as the initialization of the optimization. We employed a nonlinear minimization algorithm without constraining the phase step of the actuator elements. Details of the associated algorithm can be found in our previous work (Dou et al. 2011; Ren et al. 2012). Once the optimal phase $\Phi$ is found to generate a dark hole with a certain contrast, we compute $G_{i0}$ by using a round operation on $(\Phi/\varphi_{a})$, where $\varphi_{a}$ is the phase step of the SLM or DM. Then $G_{i0}$ will be as the start point of the optimization. In the second step, we use the global optimization solver of simulannealbnd to minimize Equation (2), by updating $G_{i}$  from $G_{i0}$ with a variation of an integer times the actuator resolution in each iteration step. The optimization converges fast with a good estimate of the start point. In the final step, we apply the optimal phase ($G_{i}*\varphi_{a}$) and compute the contrast achieved in the dark hole by using the SLM and DM, respectively.

The achievable contrast can be calculated by the empirical formula (Brown et al. 2003; Trauger et al. 2003)
\begin{equation}
\ Contrast=\pi(\frac{2\pi\sigma}{N\lambda})^{2},
\end{equation}
where $\sigma$ is the rms error in the actuator elements' surface positioning, limited by the phase step resolution; and N is the number of actuators across the pupil.

OWA will be set to $16 \lambda/D$ due to the Nyquist limit when using $32 \times 32$ actuator elements for dark hole generation. Both the SLM and DM have delivered the same performance with a contrast of $10^{-7}$. Although the DM has a relatively high phase step resolution, no contrast improvement could be achieved by performing further optimization iterations, which may indicate that the phase accuracy of the actuator elements is not the main limitation factor for dark hole generation for a clear pupil. Figure 2 shows the numerical simulation results via a $32 \times 32$ actuator DM and SLM.

According to Equation (3), further contrast improvement should be achieved by increasing the actuator numbers or reducing the size of the OWA region. Therefore in the second step, we employ a $64 \times 64$ actuator element to generate a dark hole with the same size as that using the $32 \times 32$ actuators; meanwhile, we reduce the OWA to $8 \lambda/D$, half the size of the Nyquist limit when using $32 \times 32$ actuator elements (see Figure 3 and Figure 4). Both configurations can deliver a contrast on the level of $10^{-8}$. We then test if further contrast improvements can be achieved by reducing the detection regions to $8 \lambda/D$, when using a $64 \times 64$ actuator SLM, which corresponds to a $128 \times 128$  actuator SLM to generate a dark hole with an OWA of $16 \lambda/D$ (see Figure 5). However, it is found no contrast gain can be achieved, due to the strong diffraction light for the clear pupil system.

Table 1 lists the configuration used in the simulation and associated performance, with different actuator numbers, phase accuracy, as well as dark hole size. Several conclusions can be given. (1) the actuator number, rather than the phase accuracy, is the main contrast limitation factor to generate dark hole with contrast up to $10^{-8}$. The DM with a high phase accuracy has delivered the same contrast as the SLM. (2) Contrast can be improved to $10^{-8}$ by increasing the actuator numbers, which is consistent with the result presented in Malbert' s work (1995). However, contrast cannot be improved further when the bright diffraction starlight is a dominating noise from the clear pupil, which needs to be suppressed by introducing a coronagraph. And the simulation results do not match Equation (3), when working under the photon noise dominated case due to diffraction from the telescope pupil.




\section{ HIGH-CONTRAST IMAGING FOR A CORONAGRAPH WITH WAVE-FRONT CONTROL}

 In this section, we will introduce a coronagraph to suppress diffraction light, leaving speckle noise as the dominant limitation source, which can be further corrected by the wave-front control. For demonstration purposes, we will employ the stepped-transmission filter coronagraph, although other types of coronagraph can be used. The coronagraph is composed of several stepped-transmission filters, with a finite number of steps of identical transmission in each step. The design of the coronagraph can be found in a previous paper (Ren \& Zhu, 2007) and will not be discussed here.

\subsection{Optical Configuration and Optimization Algorithm for a Coronagraph}

Figure 6 shows one of the optical configurations figures when the phase correction element (SLM or DM) is located after the coronagraph focal plane mask, the same as in our previous test to make the system compact (Liu et al. 2015). We will not show and discuss the configuration when the SLM is put before the focal plane mask, with only the simulation results presented in following sections. In the practical test, one polarizer must be inserted after L3 when using the SLM for phase-only control, which has not been shown in the figure (the configuration can be found in Liu et al. 2015). A coronagraph without wave-front control can only deliver a contrast of $10^{-7}$ in practice, due to amplitude and phase aberration, which can be interconverted by Fresnel propagation (Guyon 2005). Thus we represent the amplitude and phase aberration by using a complex function $\phi$. The aberrated pupil function of the coronagraph is then written as
\begin{equation}
\ P_{1}(u,v)=A(u,v)e^{i\phi(u,v)},
\end{equation}

where $A (u, v)$ is the perfect coronagraph pupil.

The associated focal plane complex amplitude is
\begin{equation}
\ F_{1}(x,y)=\digamma [P_{1} (u,v)]=\digamma [A(u,v)]\otimes \digamma [e^{i\phi(u,v)}] ,
\end{equation}

with $\otimes$ denoting the convolution operation; and the Fourier transformation of the stepped-transmission apodized pupil $A(u, v)$ can be found in a previous paper (Ren \& Zhu 2007).

An occulting hard-edged mask $M(x, y)$ is introduced in the focal plane to block the central bright starlight. The complex amplitude after the mask can be represented as
\begin{equation}
\ F_{2}(x,y)=F_{1}(x,y) M (x,y).
\end{equation}

Then the electric field of the optics wave on the exit pupil is the inverse Fourier transform of $F_{2}$:
\begin{equation}
\ P_{2}(u,v)=\digamma^{-1} [F_{2} (x,y)]=P_{1}(u,v) \otimes \digamma^{-1} [M(x,y)],
\end{equation}
where $\digamma^{-1}$ represents the inverse Fourier transformation.

The SLM or DM will be put after the exit pupil for wave-front control to correct the amplitude and phase aberrations, and the associated complex amplitude is then
\begin{equation}
\ P_{3}(u,v)=P_{2}(u,v)e^{i\varphi(u,v)},
\end{equation}
where $\varphi(u,v)$ is the phase provided by the SLM or DM.
The final electric field at the image plane of the camera is the Fourier transform of Equation (8):
\begin{equation}
\ F_{3}(x,y)=F_{1}(x,y) M (x,y)\otimes \digamma [e^{i\varphi(u,v)}].
\end{equation}

The associated PSF of the starlight is the square of the complex modulus of Equation (9) and we get
\begin{equation}
\ I=|F_{3}(x,y)|^{2}.
\end{equation}

In the optimization procedure, we employ the same object function of Equation (2) and the same optimization area shown in Figure (1), but with a better target contrast on the order of $10^{-10}$.

\subsection{Simulation of High-contrast Imaging for a Coronagraph}


In the simulation, we introduce a stepped-transmission filter coronagraph to suppress diffraction light from the pupil. Meanwhile, a focal occulting mask with a size of $2 \lambda/D \times 2 \lambda/D$ has been employed to further block the central bright starlight, as illustrated in Figure 6. Both optical configurations with the SLM located before and after the coronagraph focal plane mask will be presented.

\subsubsection{Configuration I: SLM located after the focal plane mask}

In this section, we will carry out numerical simulation using the SLM in a step-transmission filter based coronagraph, when putting the SLM after the coronagraph focal plane mask. We have also employed the same optical configuration in our previous high-contrast imaging tests for the step-transmission filter based coronagraph (Liu et al. 2015). Based on the current technique, we set the phase accuracy of the SLM and DM as 1/1000 and 1/10,000 waves, respectively. At first, we use both DM and SLM with $32 \times 32$  and $64 \times 64$ arrays of actuator elements. See Figure 7 and 8 for the simulation results, which show that the SLM has a relatively low phase accuracy and can deliver similar performance to the DM.

Due to the Nyquist limit, the largest size of the achievable dark hole is only $32 \lambda/D$ with current DMs in theory, which corresponds to an outer angular separation of $0.8\arcsec $(4 m class telescope at 500nm). The size of achievable dark hole can be further increased by increasing the effective number of actuators. For instance, a dark hole with an OWA of $64 \lambda/D$ should be achieved with current SLMs of $512 \times 512$ pixels working under $4 \times 4$ pixels binning mode. Future $1K \times 1K$ SLMs can correct a dark hole with an OWA of $128 \lambda/D$. Finally, we use a $90 \times 90$  pixel array SLM to generate a dark hole with an OWA extending to $45 \lambda/D $, for demonstration purposes. Simulation results are shown in Figure 9.

The summary of the simulation can be found in Table 2, which lists the different configurations of the actuator elements and detection region size, as well as the achievable contrast. It is found that the SLM with a relatively low phase accuracy can deliver a similar contrast of $10^{-10}$ to that of using a DM, when placing it after the focal mask of the step-transmission filter based coronagraph. Therefore, the phase step resolution of 1/1000 waves should be sufficient to reach the contrast level for the direct imaging of an Earth-like exoplanet in the habitable zone. Since a SLM has more controllable actuators, it can further increase the area of the high-contrast imaging region, which can be applied for the detection of mature exoplanets located at large orbital separations.

\subsubsection{Configuration II: SLM located before the focal plane mask}

We then use a different optical configuration by putting the SLM before the coronagraph focal plane mask. At first, we use a $32 \times 32$ array of actuator elements with a phase accuracy of 1/1000 waves, and set the target contrast to be $10^{-10}$. However, only a medium contrast of $4\times10^{-10}$ can be achieved. According to Equation (3), we then increase the phase accuracy and actuator element number, respectively. Figure 10 shows the achievable contrast when using a different phase accuracy of the SLM. And it is found that increasing the phase accuracy to 0.4/1000 waves should be sufficient to achieve a contrast down to $10^{-10}$. Figure 11 shows the contrast improvement achieved by increasing the actuator element number. Both simulation results are in good agreement with the Equation (3), when working under this optical configuration. And it is found that the phase accuracy of the SLM needs to be doubled to reach a contrast on the order of $10^{-10}$; however, the contrast can be further improved by increasing the actuator element number, when working under the phase accuracy of 1/1000 waves. As discussed above, SLMs possess more actuator elements, thus they could provide a competitive contrast performance compared to that by using a DM.

\section{DISCUSSION}

In this paper, we have focused on the phase quantization effect for the SLM. Other optical limitations to the SLM such as polarization stability or homogeneity of actuators will still need further analysis. In the simulation, we have used monochromatic light. The influence of polychromatic light for broadband high-contrast imaging has been discussed in the previous work (Malbert 1995; Ren \& Zhu 2011). And in a recent test, image contrasts of $3\times10^{-10}$ over $2\%$ bandwidths and $2\times10^{-9}$ over $20\%$ bandwidths have been reported (Trauger et al. 2012). More work on the polychromatic effect on SLMs in broadband high-contrast imaging is needed in the future work, since DMs have better achromatic performance than SLMs.

In practical application, the temperature stability will be one of the main contributors to affect the actual performance of both DMs and SLMs. To eliminate the influence of potential error sources, the high-contrast imaging system will be required to be put into a vacuum tank with precise temperature control. And $50\%$ light will be reduced when SLMs work under the phase-only mode by using a polarizer. To fully use the $100\%$ of the incoming light, one solution is to employ the configuration shown in a previous work (Ren \& Zhu 2011), in which one polarized prism splits the light into two beams with each beam of the same layout, and two identical light beams are finally combined in the focal plane imaging camera; the other solution is to use the polarization-independent SLM developed by Meadowlark Optics.

Another potential way to improve the performance of SLMs is by increasing the current phase step of electronics by a factor of 2 or better, which can improve the phase accuracy of the SLM accordingly to meet the contrast requirement for detection of Earth-like planets in the habitable zone of a solar-type star.

\section{CONCLUSIONS AND FUTURE WORK}

In this paper, we present a feasibility study of using a SLM for wave-front control in extreme high contrast imaging. The dark hole optimization algorithm and associated numerical simulation have been conducted for both a clear pupil and the coronagraph system. It is found that the actuator number is the main contrast limitation factor when used for a clear pupil system. Contrast can be improved from $10^{-7}$ to $10^{-8}$ by increasing the actuator number or reducing the dark hole size. However, no further contrast gain can be achieved when increasing the actuator number, due to diffraction noise being dominant for a clear system with a contrast limitation up to $10^{-8}$. For the coronagraph system, we have employed two optical configurations by putting the SLM before and after the focal plane mask. Several conclusions can be drawn from the simulation, shown as follows: (1) In optical configuration I (SLM located after the focal plane mask), the SLM with a relatively low phase accuracy (1/1000 waves of rms) can provide a similar performance to a DM and should be sufficient to reach a high contrast level of $10^{-10}$. (2) In optical configuration II (SLM located before the focal plane mask), we need to increase the phase accuracy to 0.4/1000 waves or increase the actuator element number accordingly to reach a contrast on the order of $10^{-10}$ and the simulation results matched the Equation (3). (3) A large size of dark hole can be achieved via SLMs with more controllable actuator elements, and it could provide a competitive contrast performance to DMs.

In our future work, we will analyze and discuss the chromaticity and the other optical limitations of the SLM elements, including the polarization quality, stability, homogeneity with each actuator element, etc.




\acknowledgments
We thank the anonymous referee for valuable comments, which have significantly improved the manuscript. This work was supported by the NSFC (Grant Nos. 11433007, 11661161011, 11220101001, 11328302, 11373005, and 11303064), the ¡±Strategic Priority Research Program¡± of the Chinese Academy of Sciences (Grant No. XDA04075200), the International Partnership Program of Chinese Academy of Sciences (Grant No.114A32KYSB20160018), as well as the special fund for astronomy of CAS (2015-2016). Part of the work described in this paper was carried out at California State University Northridge, with the support from Mt. Cuba Astronomical Foundation.

\clearpage



\begin{figure}
\epsscale{0.6}
\plotone{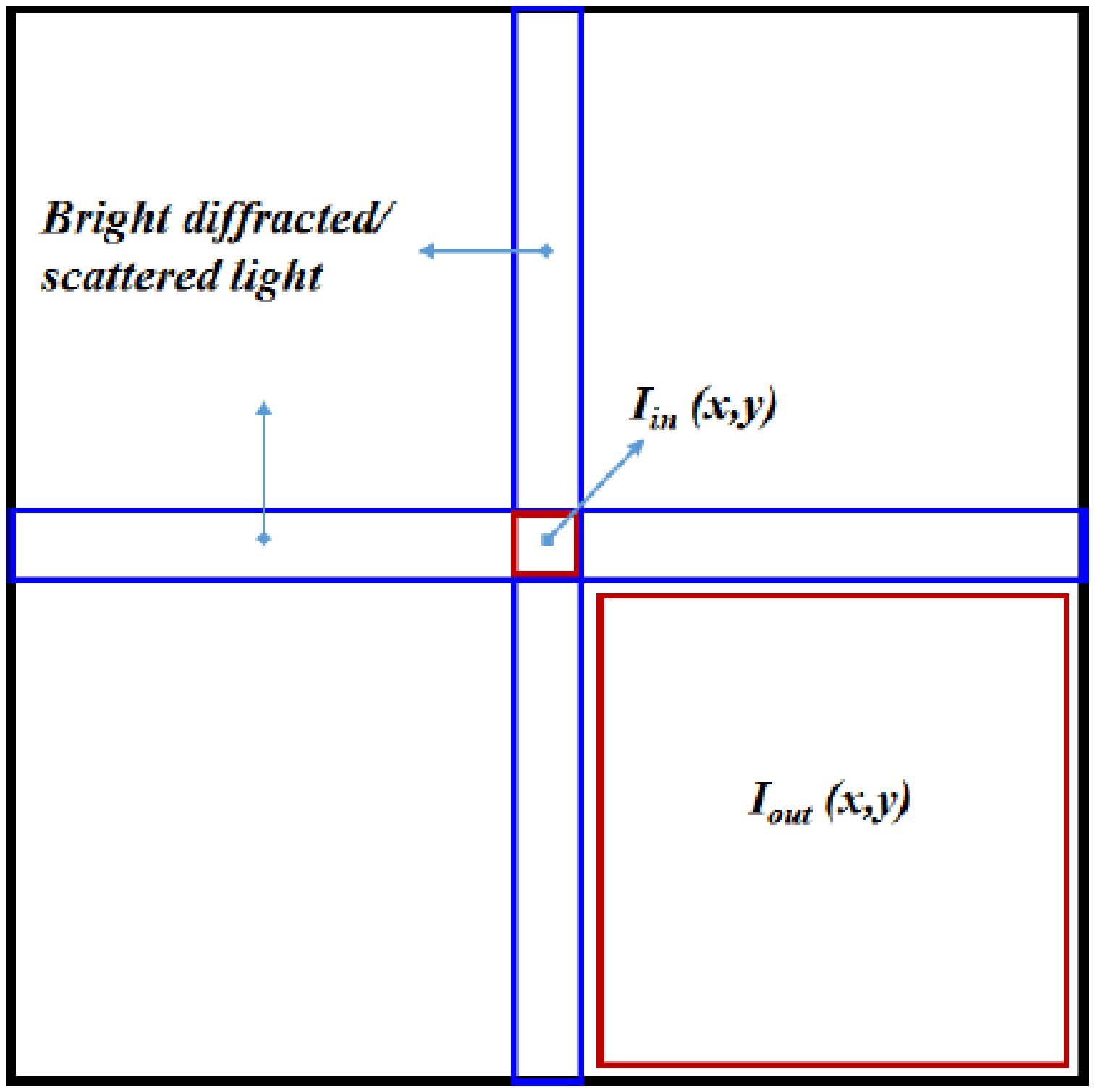}
\caption{Definition of the optimization Area.}
\end{figure}


\begin{figure}
     \begin{center}
        \subfigure{%
            \label{fig:first}
            \includegraphics[width=0.5\textwidth]{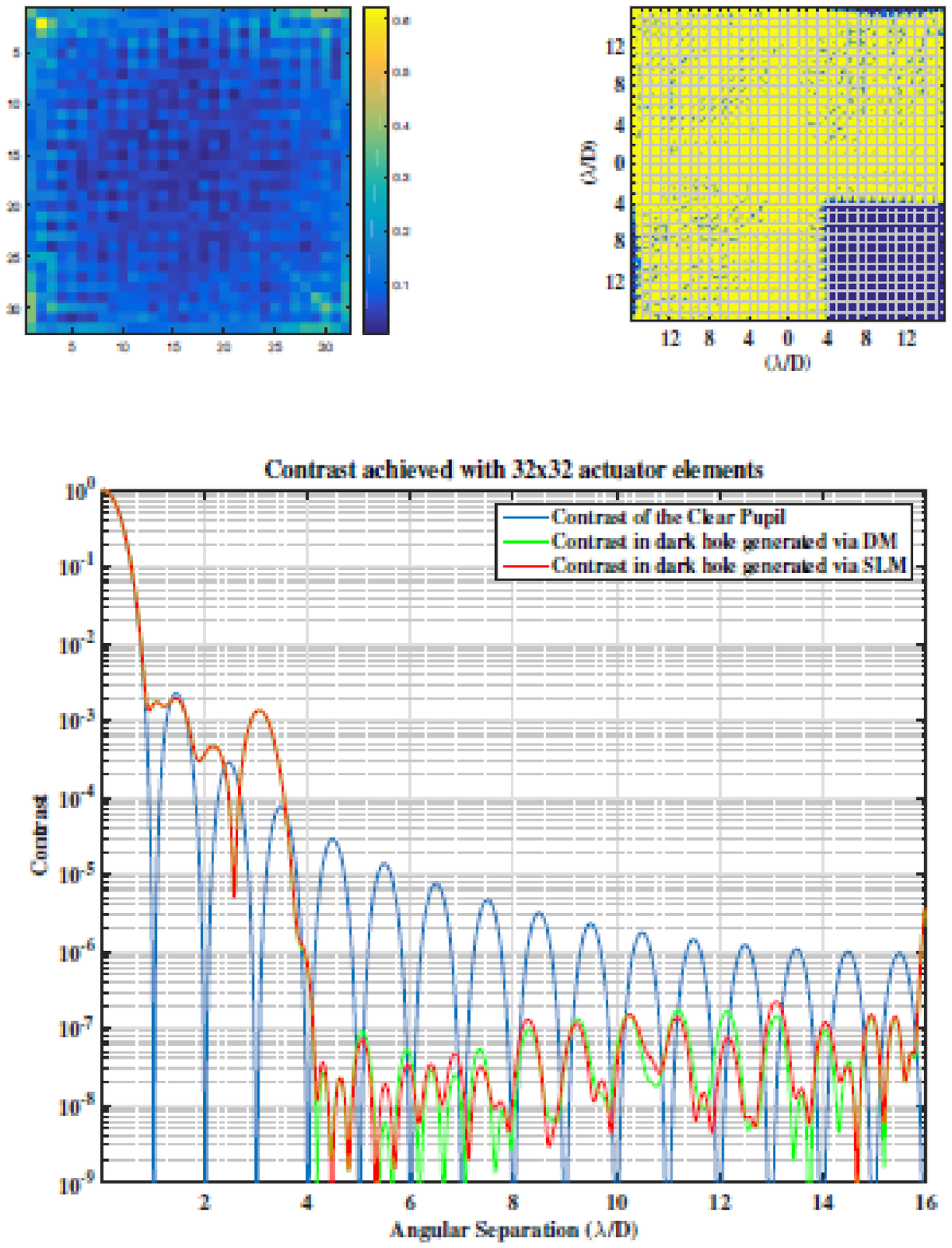}
        }%

    \end{center}
    \caption{%
        Dark hole generation for a clear pupil via a $32 \times 32$ actuator DM and SLM: phase map in the unit of working wavelength (upper left), the associated PSF image via SLM (upper right) and the achieved contrast curve by using the SLM and DM, with associated phase accuracy (bottom).
     }%
   \label{Fig3}
\end{figure}

\begin{figure}
     \begin{center}
        \subfigure{%
            \label{fig:first}
            \includegraphics[width=0.5\textwidth]{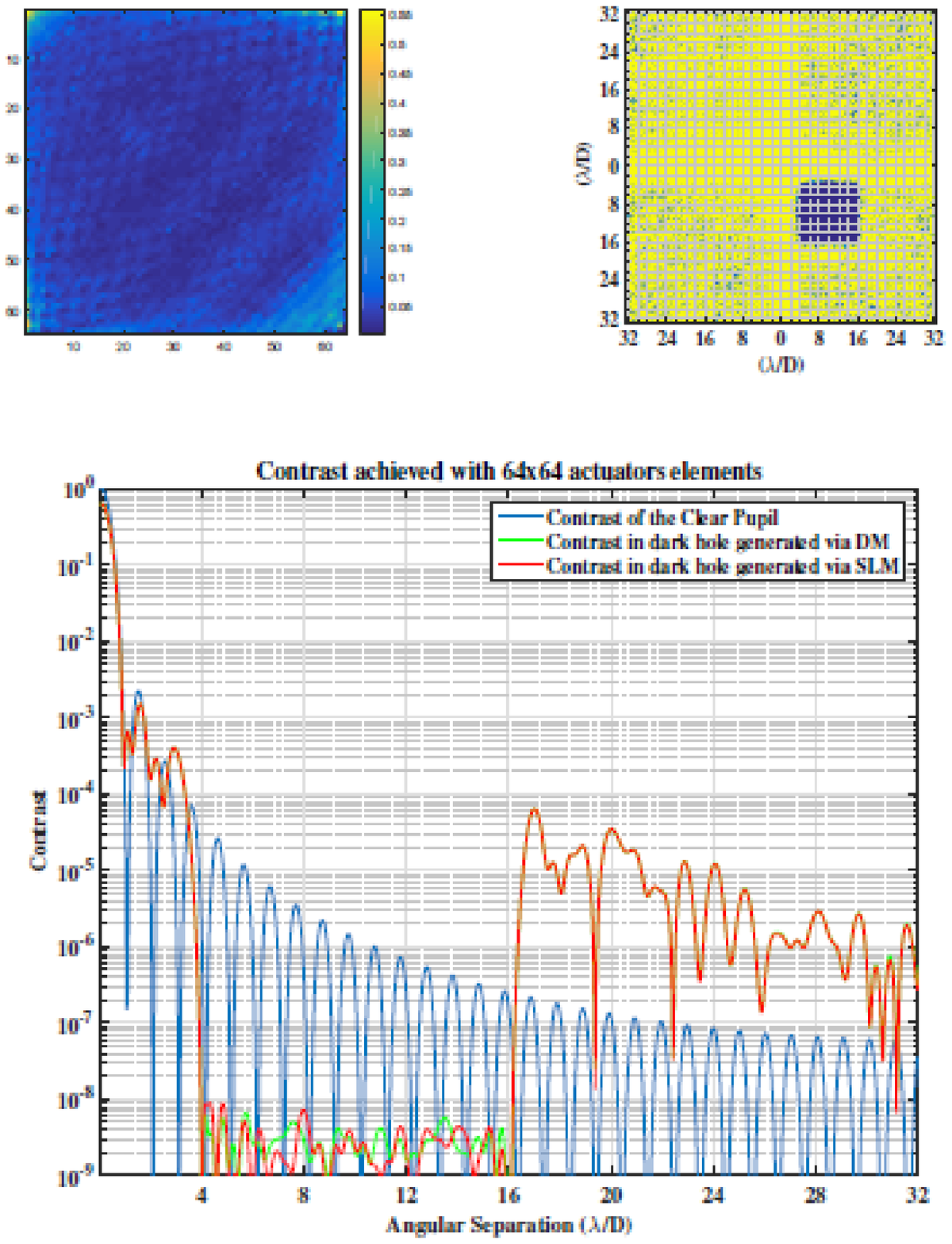}
        }%

    \end{center}
    \caption{%
        Dark hole generation for a clear pupil via a $64 \times 64$ actuator DM and SLM, but with a small OWA ($16 \lambda/D$): phase map in the unit of working wavelength (upper left), the associated PSF image via the SLM (upper right) and the achieved contrast curve by using the SLM and DM, with associated phase accuracy (bottom).
     }%
   \label{Fig4}
\end{figure}

\begin{figure}
     \begin{center}
        \subfigure{%
            \label{fig:first}
            \includegraphics[width=0.5\textwidth]{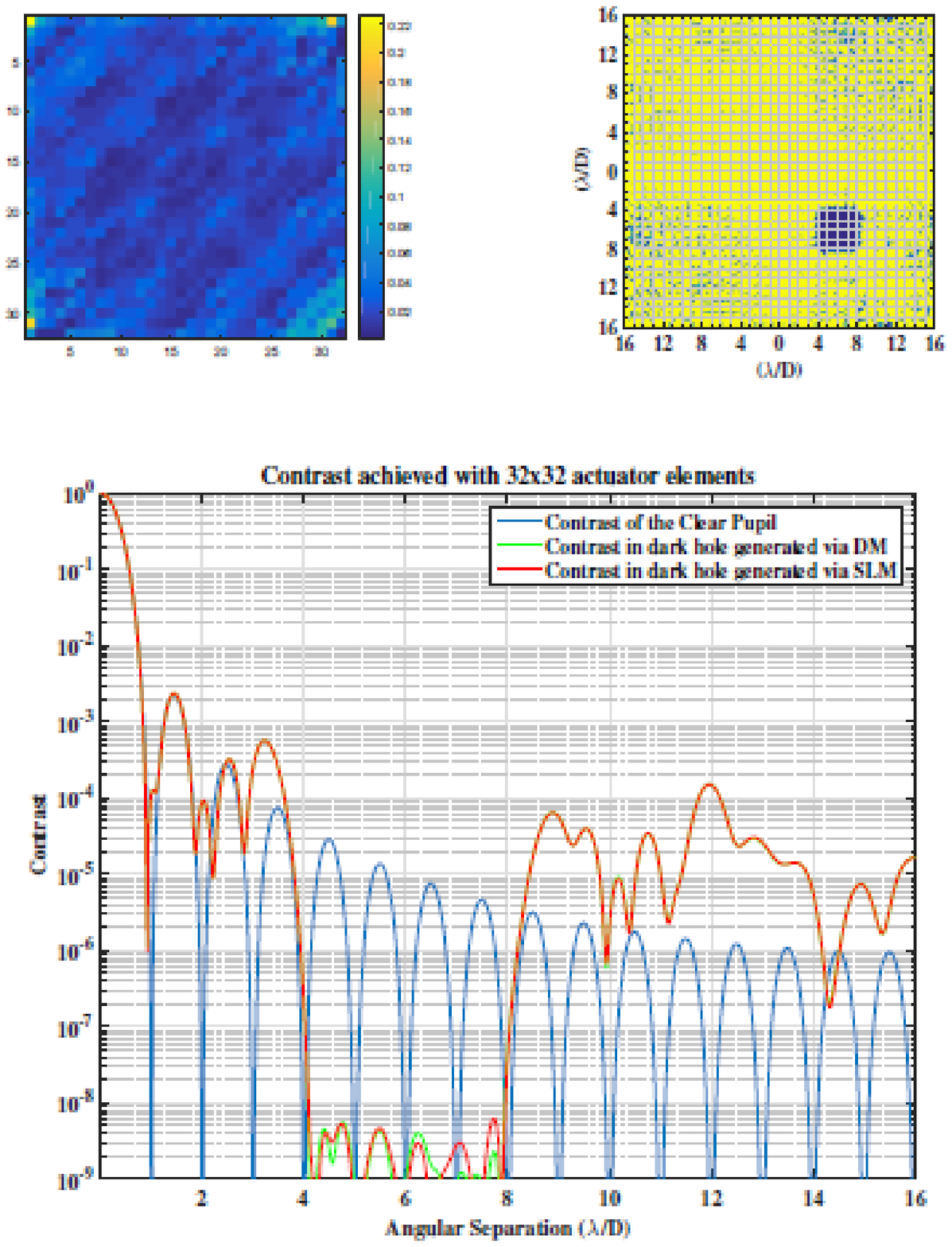}
        }%

    \end{center}
    \caption{%
        Dark hole generation for a clear pupil via a $32 \times 32$ actuator DM and SLM, but with a small OWA ($8 \lambda/D$): phase map in the unit of working wavelength (upper left), the associated PSF image via the SLM (upper right) and the achieved contrast curve by using the SLM and DM, with associated phase accuracy (bottom).
     }%
   \label{Fig5}
\end{figure}

\begin{figure}
     \begin{center}
        \subfigure{%
            \label{fig:first}
            \includegraphics[width=0.5\textwidth]{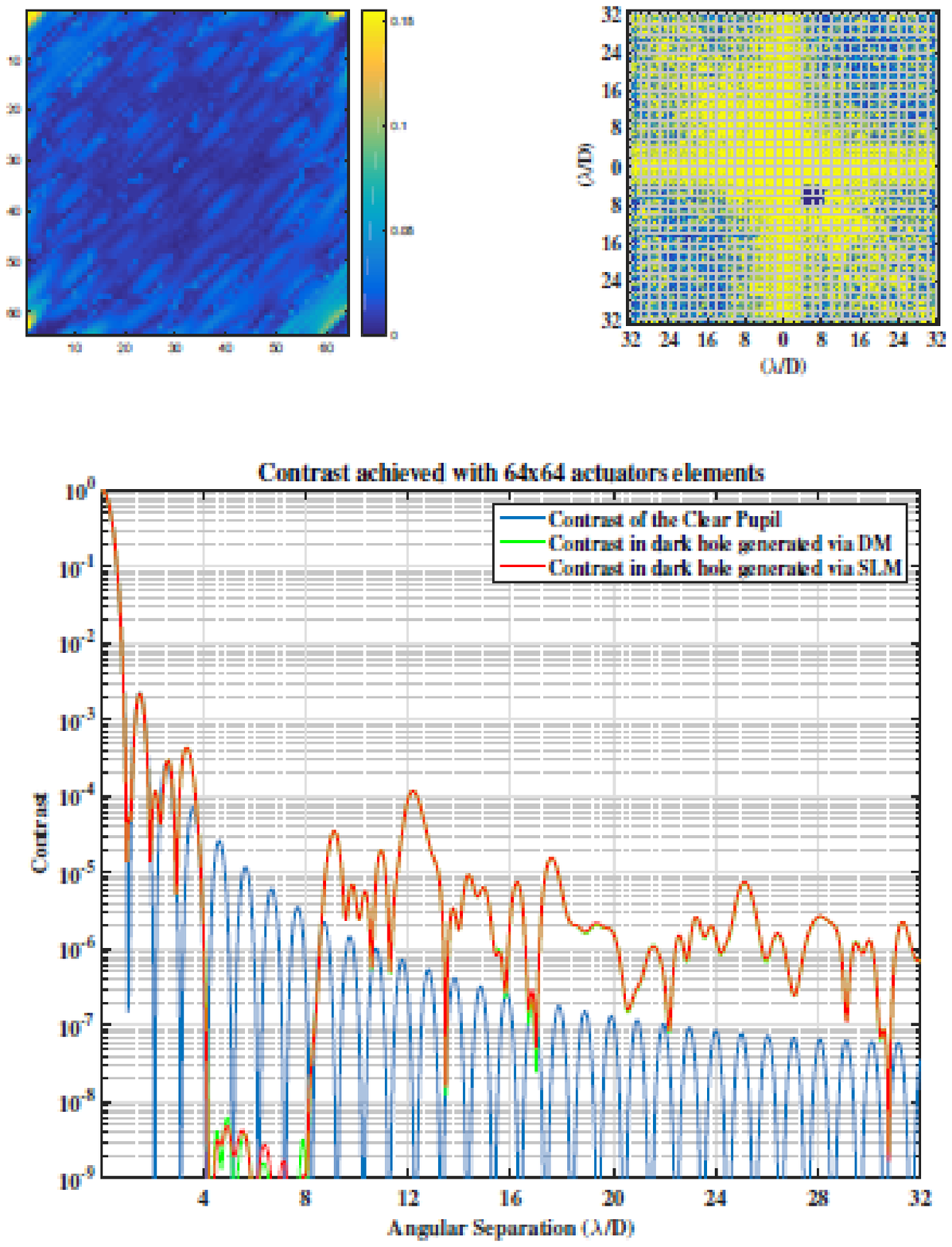}
        }%

    \end{center}
    \caption{%
        Dark hole generation for a clear pupil via a $64 \times 64$ actuator DM and SLM, but with a small OWA ($8 \lambda/D$): phase map in the unit of working wavelength (upper left), the associated PSF image via the SLM (upper right) and the achieved contrast curve by using the SLM and DM, with associated phase accuracy (bottom).
     }%
   \label{Fig6}
\end{figure}

\begin{figure}
\epsscale{1.0}
\plotone{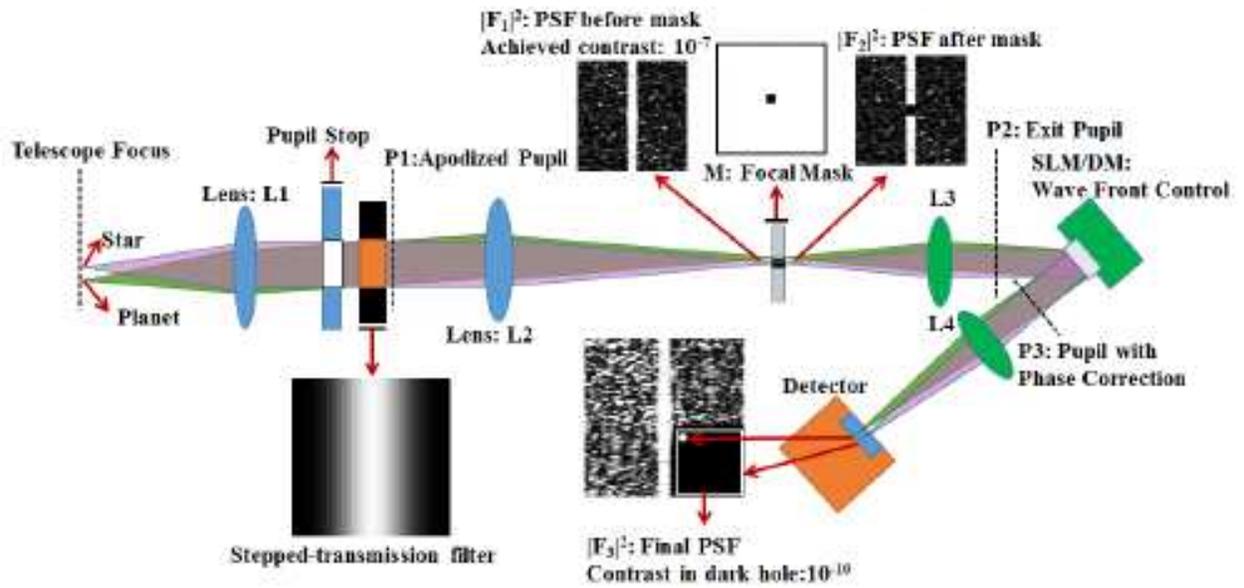}
\caption{Optics layout of the corongraph with a focal occulting mask and phase corrections.\label{fig6}}
\end{figure}
\begin{figure}

     \begin{center}
        \subfigure{%
            \label{fig:first}
            \includegraphics[width=0.5\textwidth]{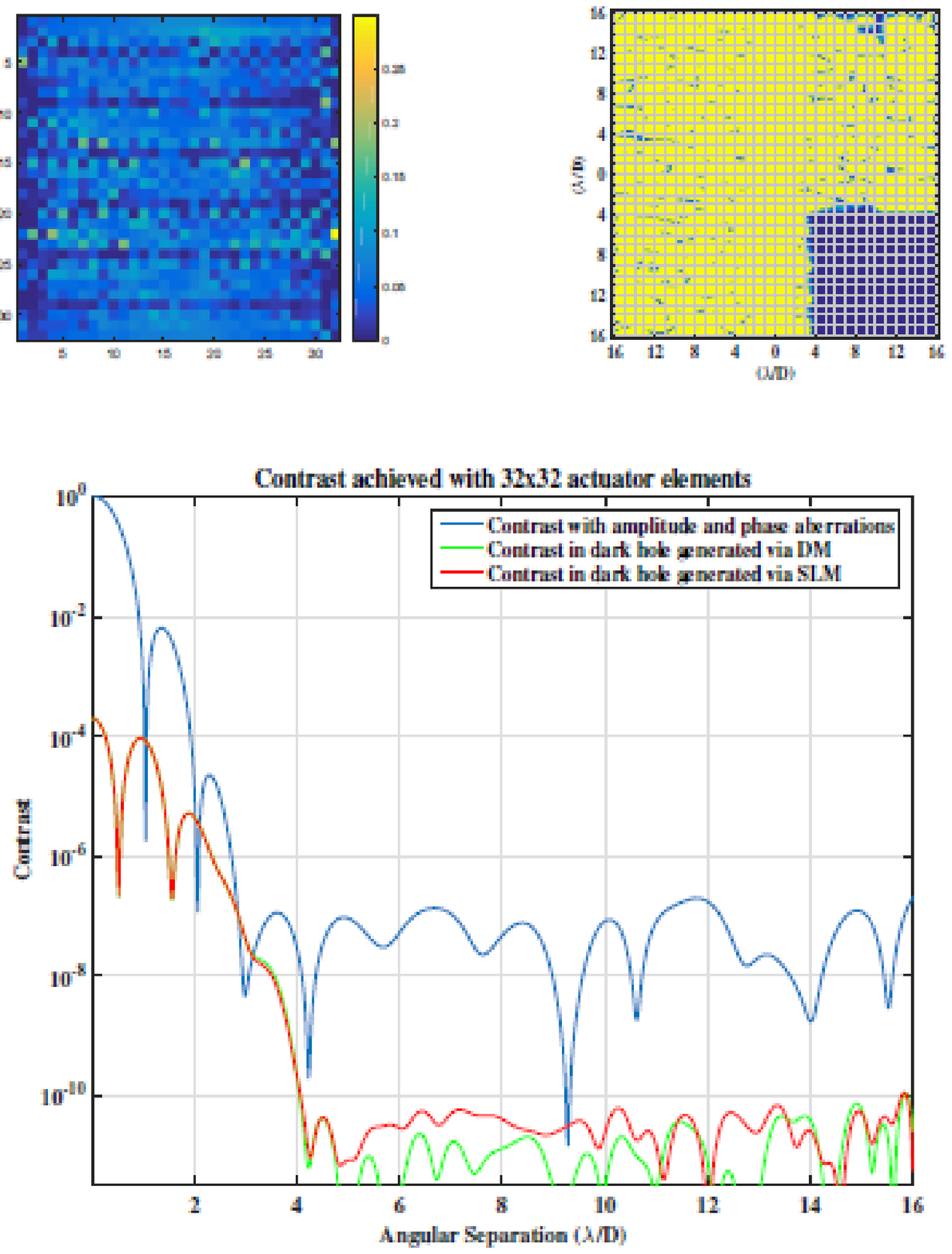}
        }%

    \end{center}
    \caption{%
        Dark hole generation for a coronagraph via a $32 \times 32$ actuator DM and SLM: phase map in the unit of working wavelength (upper left), the associated PSF image via the SLM (upper right) and the achieved contrast curve by using the SLM and DM, with associated phase accuracy (bottom).
     }%
   \label{Fig7}
\end{figure}

\begin{figure}
     \begin{center}
        \subfigure{%
            \label{fig:first}
            \includegraphics[width=0.5\textwidth]{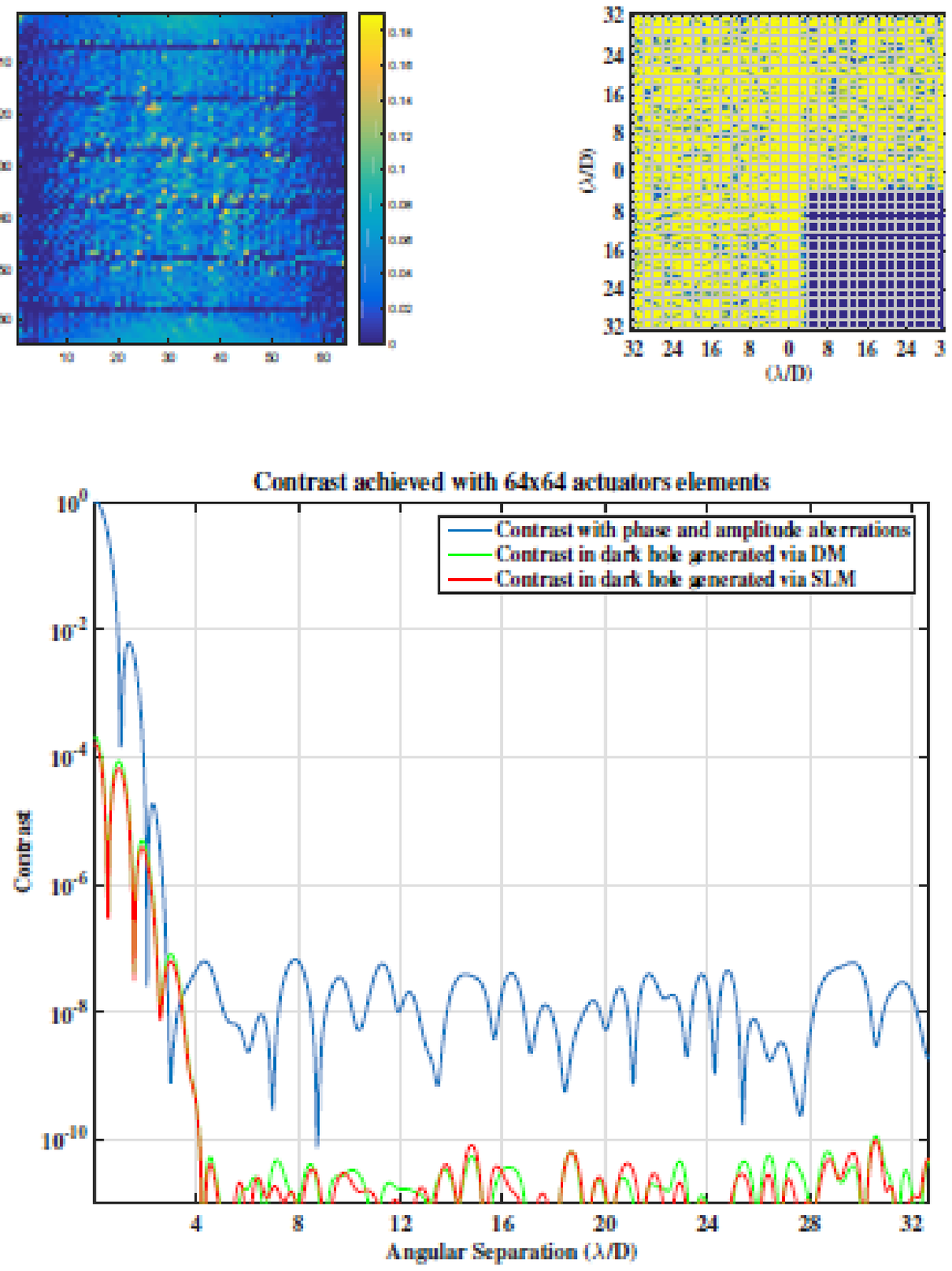}
        }%

    \end{center}
    \caption{%
        Dark hole generation for a coronagraph via a $64 \times 64$ actuator DM and SLM: phase map in the unit of working wavelength (upper left), the associated PSF image via the SLM (upper right) and the achieved contrast curve by using the SLM and DM, with associated phase accuracy (bottom).
     }%
   \label{Fig8}
\end{figure}

\begin{figure}
     \begin{center}
        \subfigure{%
            \label{fig:first}
            \includegraphics[width=0.5\textwidth]{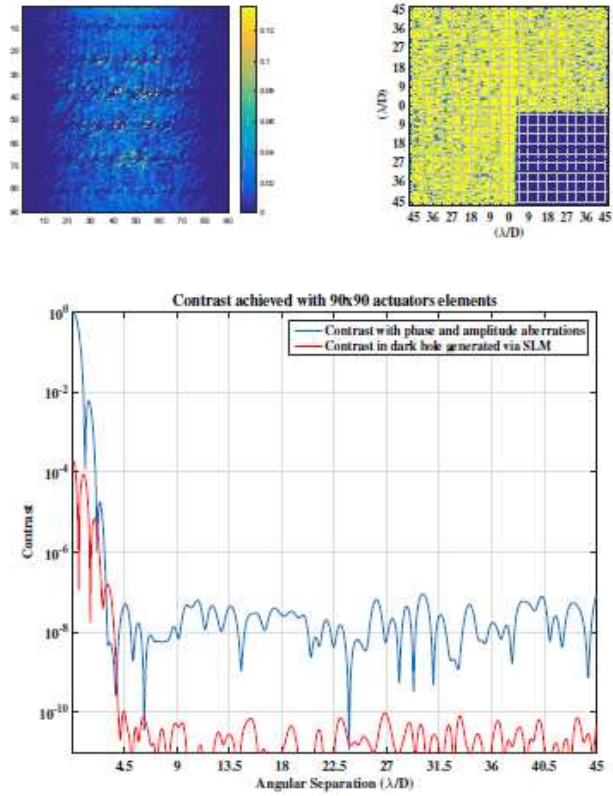}
        }%

    \end{center}
    \caption{%
        Dark hole generation for a coronagraph via a $90 \times 90$ actuator SLM: phase map in the unit of working wavelength (upper left), the associated PSF image via the SLM (upper right) and the achieved contrast curve by using the SLM (bottom).
     }%
   \label{Fig9}
\end{figure}

\begin{figure}
     \begin{center}
        \subfigure{%
            \label{fig:first}
            \includegraphics[width=0.5\textwidth]{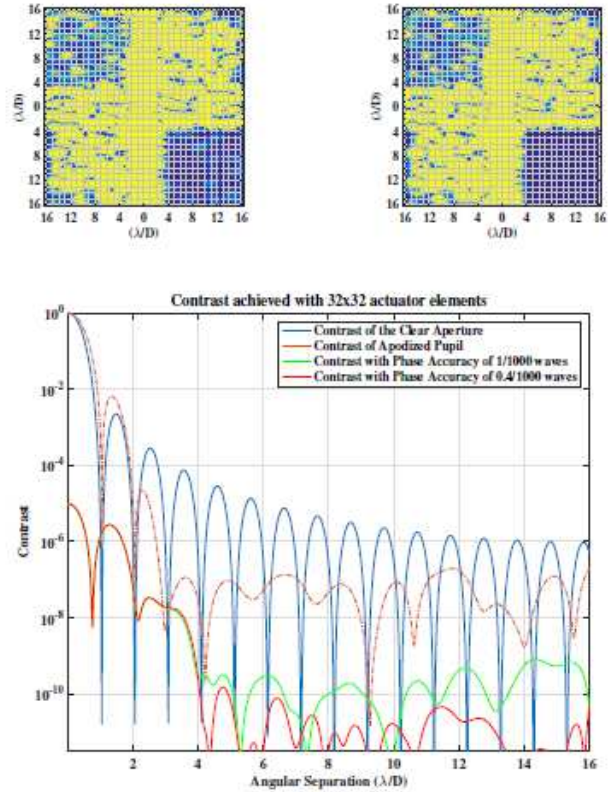}
        }%

    \end{center}
    \caption{%
        Dark hole generation for a coronagraph via a $32 \times 32$ actuator SLM: PSF image via the SLM with phase accuracy of 1/1000 waves (upper left), PSF image via the SLM with phase accuracy of 0.4/1000 waves (upper right) and the achieved contrast with different phase accuracy (bottom).
     }%
   \label{Fig10}
\end{figure}

\begin{figure}
     \begin{center}
        \subfigure{%
            \label{fig:first}
            \includegraphics[width=0.5\textwidth]{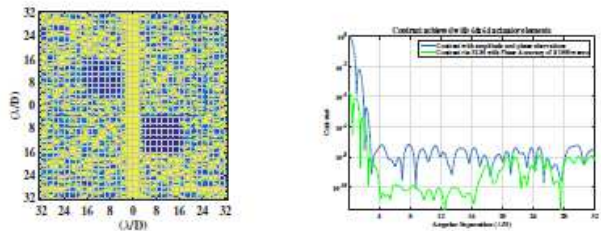}
        }%

    \end{center}
    \caption{%
        Dark hole generation for a coronagraph via a $64 \times 64$ actuator SLM: PSF image via the SLM with phase accuracy of 1/1000 waves (left) and the achieved contrast (right).
     }%
   \label{Fig11}
\end{figure}

\clearpage
\begin{deluxetable}{lccccc}

\tabletypesize{\scriptsize}
\tablecaption{Configuration and result of dark hole simulation for the clear pupil system.\label{tbl-1}}
\tablewidth{0pt}
\tablehead{
\colhead{Figures}           & \colhead{Actuator number}      &
\colhead{OWA($\lambda/D$)}& \multicolumn{2}{c}{Achievable Contrast(gain)}\\

\cline{4-5}
 & &            & \multicolumn{1}{c}{SLM(accuracy: $\lambda/1000$)}  &
\multicolumn{1}{c}{DM(accuracy: $\lambda/10,000$)} }
\startdata
FIG.02 & $32 \times 32$ & 16 & $10^{-7} (100)$ & $10^{-7} (100)$  \\
FIG.03 & $32 \times 32$ & 8 & $10^{-8} (1000)$ & $10^{-8} (1000)$  \\
FIG.04 & $64 \times 64$ & 16 & $10^{-8} (1000)$ & $10^{-8} (1000)$  \\
FIG.05 & $64 \times 64$ & 8 & $10^{-8} (1000)$ & $10^{-8} (1000)$  \\
\enddata
\end{deluxetable}

\begin{deluxetable}{lccccc}

\tabletypesize{\scriptsize}
\tablecaption{Configuration and result of dark hole simulation for a coronagraph.\label{tbl-2}}
\tablewidth{0pt}
\tablehead{
\colhead{Figures}           & \colhead{Actuator number}      &
\colhead{OWA($\lambda/D$)}& \multicolumn{2}{c}{Achievable contrast}\\

\cline{4-5}
 & &            & \multicolumn{1}{c}{SLM(accuracy: $\lambda/1000$)}  &
\multicolumn{1}{c}{DM(accuracy: $\lambda/10,000$)}           }
\startdata
FIG.07 & $32 \times 32$ & 16 & $10^{-10}$ & $10^{-10}$ \\
FIG.08 & $64 \times 64$ & 32 & $10^{-10}$ & $10^{-10}$ \\
FIG.09 & $90 \times 90$ & 45 & $10^{-10}$ & $\sim$ \\
\enddata
\end{deluxetable}
%





\end{document}